# Manifold partitioning induced sequential optical reasoning and decision framework for photonic computing

Zhihao Li[1]†, Jing Pan[1]†*, Wei Yan[1, 2], Yu Xie[2], Lingmei Ma[2], Xiaoyu Sun[1], Min Qiu[1, 2]*

1. Zhejiang Key Laboratory of 3D Micro/Nano Fabrication and Characterization, Department of Electronic and Information Engineering, School of Engineering, Westlake University, Hangzhou, Zhejiang 310030, China.

2. Westlake Institute for Optoelectronics, Fuyang, Hangzhou, Zhejiang 311400, China.

*Corresponding author, E-mail: panjing75@westlake.edu.cn, qiumin@westlake.edu.cn

†These authors contributed equally to this work.

## Abstract

Real-world data are intrinsically embedded in highly entangled manifolds, making the extraction of separable representations a central challenge for artificial intelligent (AI) systems. While optical neural networks (ONNs) offer ultrafast and energy-efficient data processing, their capacity is constrained by limited physical depth. Here, we introduce a sequential optical reasoning and decision (SORD) framework, an architecture that performs time-sequenced hierarchical inference by decomposing global tasks into coarse-to-fine steps via geometry-guided data partitioning. At each step, SORD executes small reasoning via dynamic operator selection, effectively reducing the overall task complexity without scaling up physical architecture. Experimentally, SORD enables a single-layer diffractive ONN to achieve otherwise intractable 100-class optical fiber speckle classification with 94% accuracy and a system energy efficiency of 23.3 TOPS/W. This high-fidelity recognition is further examined in a human-machine interface, featuring real-time interactive all-optical sensing. Overall, our work establishes a scalable and hardware-efficient approach to expanding the effective expressivity of compact photonic AI systems, and may advance their deployment in applications requiring real-time sensing, inference, and control.

## Introduction

Physical data typically consist of interdependent variables that form highly entangled manifolds rather than isolated clusters[1, 2]. Due to their nonlinear correlations and overlapping class distributions, these data structures complicate the direct extraction of latent representations compared with the low-dimensional form[3]. While intelligent systems such as deep neural networks[4] require efficient feature extraction to process raw sensory inputs, disentangling such complex data spaces imposes a heavy computational burden, often requiring significant network depth and substantial computational power[5, 6].

Optical neural networks[7, 8, 9, 10, 11, 12] (ONNs) have emerged as a promising paradigm for high-performance computing by mapping neural computations onto the intrinsic degrees of freedom of optical fields. By exploiting the analog nature of light, ONNs enable parallel[13, 14, 15, 16], low-latency data execution[17], demonstrating remarkable success in tackling computationally intensive tasks such as high-speed image classification[18], multidimensional optical field processing[19], solving NP-hard problems[20], and complex partial differential equations[21] at the speed of light. Recent demonstrations have reported energy efficiencies exceeding $10^2$ TOPS/W in large-scale diffractive systems and sub-nanosecond inference latency in integrated photonic circuits, highlighting their potential for real-time edge intelligence[22].

However, conventional ONNs are inherently limited when handling tasks that require high representational capacity due to their shallow effective depth[23, 24, 25]. In the absence of optical nonlinearities, cascaded diffractive layers collapse into an equivalent linear transformation, which imposes a practical ceiling on system performance: increasing the number of layers or neurons yields diminishing returns once the linear solution space is saturated.[26]

In recent years, researchers have sought to introduce optical nonlinearities (e.g., nonlinear materials[27], atomic media[24], Kerr effects[28], and carrier-induced nonlinearities[29]), employ optoelectronic (O–E–O) schemes[30], and adopt actively tuned photonics (e.g., Mach–Zehnder interferometer meshes[31] and micro-resonator-based processors[32]) to break the linear degeneracy. While these approaches expand the computational capability of ONNs, they inevitably introduce increased energy consumption and system complexity, thereby compromising the intrinsic advantages of passive optical computing.

Alternatively, temporal multiplexing provides another pathway to scale the computational depth of passive optical architectures without sacrificing energy efficiency. Recent studies have shown that recurrent scattering[33, 34, 35], time-sequential propagation[36], and reconfigurable diffractive processing[37] can substantially expand the functional capacity of linear optical hardware by effectively stacking multiple transformations in the temporal domain. While these approaches effectively deepen ONNs, their processing sequences are generally fixed and predetermined. Building on this foundation, whether the system can perform conditional computation remains to be investigated.

In this work, we propose a sequential optical reasoning and decision (SORD) framework that transposes the algorithmic principle of decision trees into the optical domain, formulating inference as a state-dependent, hierarchical reasoning process. By decomposing global tasks into a sequence of coarse-to-fine steps, SORD reduces the overall task complexity by focusing on localized sub-tasks, enabling the effective resolution of complex classes. This paradigm enhances representational capacity without proportional increases in architectural complexity or nonlinear components, thereby alleviating the dependence on hardware scaling. We experimentally validate the efficacy of SORD framework by achieving a 100-class optical fiber speckle classification with 94% accuracy and a system energy efficiency of 23.3 TOPS/W, and further implement a robust human-machine interface that facilitates real-time interactive control driven entirely by in-situ optical sensing. By shifting complexity from structural depth to sequential reasoning, SORD establishes a new paradigm for achieving scalable, high-capacity information processing in compact photonic systems.

## Information-theoretic and geometric mechanisms of the SORD

The performance of ONNs is ultimately determined by their ability to approximate a target mapping, which can be quantified by the approximation error ($\varepsilon_{\mathrm{approx}}$). A smaller $\varepsilon_{\mathrm{approx}}$ indicates higher fidelity in function realization and thus improved computational performance, establishing a fundamental criterion that directly governs the design and scaling of optical neural architectures. The approximation error $\varepsilon_{\mathrm{approx}}$ is fundamentally bounded and follows a power-law scaling[38] governed by a dimensionless information density ratio $C_{task}/D_{sol}$, which captures the balance between the inherent task complexity $C_{\mathrm{task}}$ and the solution space dimensionality of the system $D_{\mathrm{sol}}$[26, 39, 40]:

$$\varepsilon_{approx} = A\left(\frac{C_{task}}{D_{sol}}\right)^{s/d} = A\left(\frac{log_2(C) + H(X|Y)}{D_{sol}}\right)^{s/d} \tag{1}$$

where $A$ is an empirical coefficient determined by system calibration, $s$ represent the smoothness of the target fitting function, and $d$ denote the effective dimensions of the data manifold. From an information-theoretic perspective[41], $C_{\mathrm{task}}$ can be decomposed into the label entropy $H(Y)\approx\log_2(C)$ and the conditional entropy $H(X|Y)$ [40], capturing inter-class uncertainty and intra-class variability, respectively (Supplementary Information Note 1).

This scaling relation suggests two possible routes to reduce the approximation error: increasing the solution space dimensionality $D_{\mathrm{sol}}$, or reducing the intrinsic task complexity $C_{\mathrm{task}}$. Expanding $D_{\mathrm{sol}}$ relies on increasing physical degrees of freedom, yet is fundamentally constrained by the space–bandwidth product, fabrication complexity, and noise accumulation[42, 43, 44]. As further scaling of $D_{\mathrm{sol}}$ faces diminishing returns, attention naturally shifts to reducing the effective task complexity $C_{\mathrm{task}}$: since the label entropy $H(Y)\approx\log_2(C)$ is fixed, reducing $C_{\mathrm{task}}$ requires suppressing the conditional entropy $H(X|Y)$. Here, we adopt the Fisher discriminant ratio of inter-class distance $d_{\mathrm{inter}}$ to intra-class variation $d_{\mathrm{intra}}$ as a geometric proxy[45, 46] to approximate $H(X|Y)$ variation (Supplementary Information Note 2). From a geometric perspective, highly entangled tasks exhibit a low Fisher discriminant ratio ($d_{\mathrm{inter}}/d_{\mathrm{intra}}\sim\mathcal{O}(1)$), where inter-class separation is comparable to intra-class variation, leading to strong distributional overlap and high conditional entropy $H(X|Y)$[47, 48]. Importantly, this limitation does not stem from insufficient training or optimization of the ONN, but from the topological incompatibility between the entangled data manifold and the linear solution space[9, 24, 49]. Dictated by topological invariance, linear transformations act as homeomorphic mappings that can only deform but do not alter the underlying topology of the feature space.

Here, we introduce a sequential optical reasoning and decision (SORD) framework, which introduces a time-multiplexed, hierarchical reasoning paradigm (Fig. 1a). The overall mapping $F_{\text{SORD}}$ takes the form:

$$y_k = W_k x, \qquad W_k = I(d_{k-1}), k = 1, \ldots, K \tag{2}$$

where $y_k$ denotes the stage-$k$ response of the input $x$ under the activated operator $W_k$ with $I(\cdot)$ serves as a dynamic routing gate that activates the $W_k$ conditional on the prior decision state $d_{k-1}$. The final SORD output is determined by the final decision state $d_K$ after $K$ sequential reasoning steps.

By utilizing time-multiplexed diffractive operators ($W_1$, $W_2$, $\ldots W_K$), SORD decomposes complex global inference into a sequence of temporally ordered reasoning steps. As a result, the effective discriminative ratio increases after $K$ iterations, corresponding to a progressive reduction in the conditional entropy $H(X|Y)$[50] and thereby simplifying the effective task complexity $C_{\text{task}}$. (Supplementary Information Note 2)

Operationally, SORD process a static optical input through a time-sequenced cascade of diffractive operates as Fig. 1b demonstrates. First, the input passes through an initial, pre-trained optical operator $W_1$ (the "optical reasoning" step), which yields an intermediate output $result_1$. Based strictly on this preliminary outcome, the system makes a logical "decision" to choose and activate the most appropriate subsequent optical operator from a trained operator library. This process continues until the features are well separated into their corresponding classes.

At the hardware level (Fig. 1c), SORD avoids complex physical layering by utilizing a scalable, time-multiplexed optical architecture. High-dimensional field transformations are implemented by a SLM, which dynamically loads phase masks at discrete temporal steps ($t_1$, $t_2$, $t_3$) under the control of the SORD decision process, enabling hierarchical inference without altering the physical input.

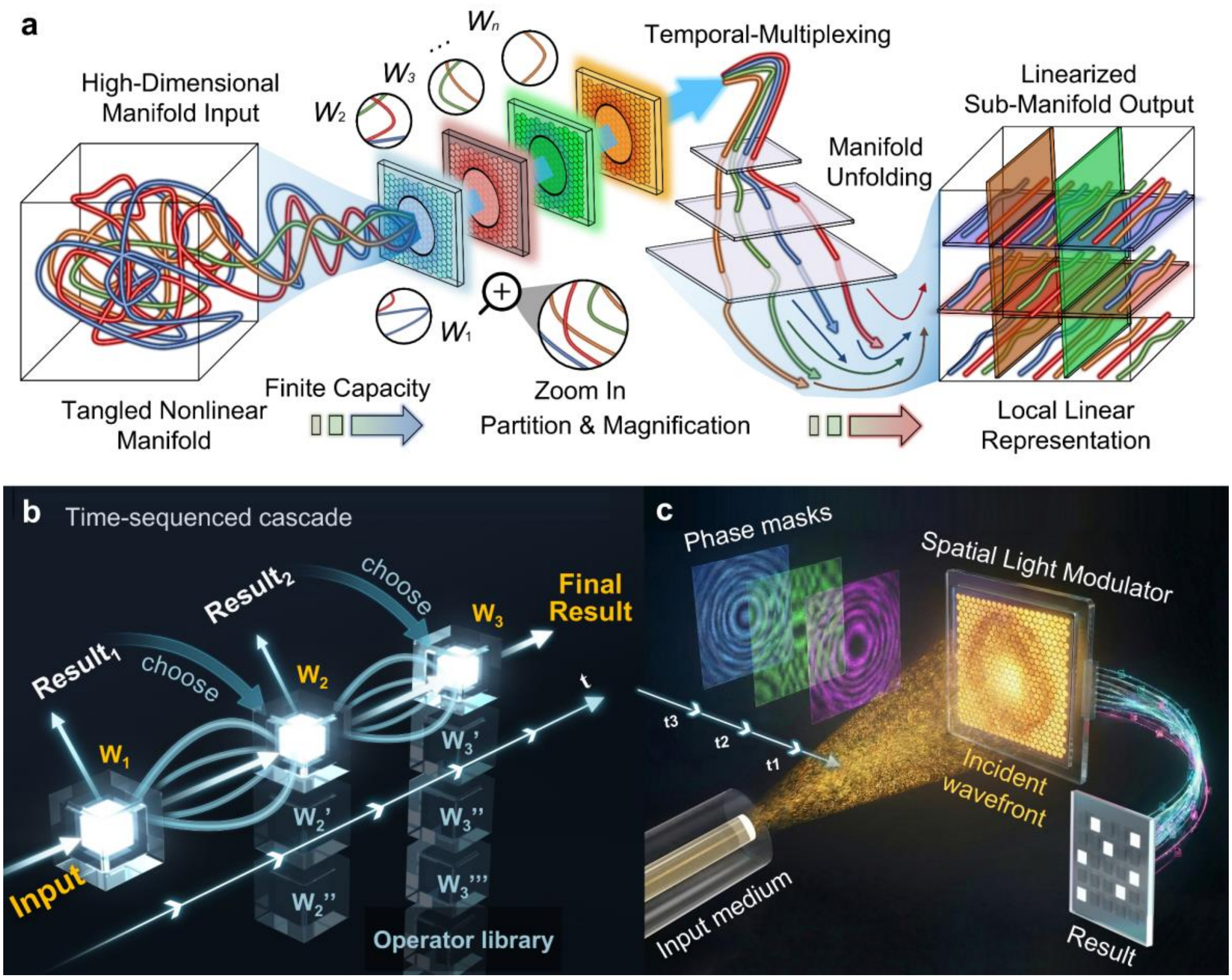

**Fig. 1 | Overview of the SORD framework.** **a**, Conceptual illustration of the SORD manifold rectification mechanism via operators $W_i$. **b**, Logical architecture of the sequential reasoning process with conditional branching. **c**, Physical implementation of the SORD paradigm using a time-multiplexed phase modulation on an SLM.

## Geometry-guided class partitioning for SORD

The performance of SORD is influenced by the underlying class partitioning, as it determines how the feature space is decomposed into sub-manifolds and thereby controls the residual conditional complexity at each stage. To minimize this complexity, the partitioning strategy must respect the natural affinities between classes rather than imposing arbitrary splits[51]. SORD adopts a geometry-guided partitioning approach by computing the Euclidean distances between class centroids $d_{i,j}$ (Supplementary Information Note 3) and grouping classes according to a minimum-distance criterion.

We evaluate the efficiency of this geometry-guided class partitioning on standard computer vision benchmarks: MNIST, Fashion-MNIST, and CIFAR-10. The t-distributed stochastic neighbor embedding (t-SNE) visualization in Fig. 2a reveals the target classes present densely clustered, highly entangled non-convex manifolds[52]. This entanglement is further quantified using Fisher criterion (Supplementary Information Note 3) to generate pairwise class-similarity heatmaps (Fig. 2b), which exhibit block-diagonal structures with high-affinity clusters (dark red blocks) indicating severely overlapping class boundaries.

To test the geometric partitioning hypothesis, we partition these datasets using the predefined strategy; compared to random splitting, this structured division yields a substantial improvement in linear separability. We physically verify this rectification by estimating the probability density functions of intra-group $P_{\text{intra}}$ versus inter-group $P_{\text{inter}}$ feature distances and quantifying their overlap area $OA = \int \min\,(P_{\text{intra}}(x), P_{\text{inter}}(x)dx$ (Supplementary Information Note 4). As shown in Fig. 2c, the geometry-guided class partitioning effectively resolves the inherent ambiguity, leading to a reduction in OA that implies a lower conditional entropy $H(X|Y)$. Under a binary classification (K = 2), the OA decreases from 97% to 73% for MNIST, from 76% to 64% for Fashion-MNIST, and from 99% to 90% for CIFAR-10. Moreover, as the number of classes increases (K = 3, 4), the geometry-guided partitioning remains effective, with consistently lower OA values than those obtained from random partitioning under the same conditions.

This improved separability directly translates into enhanced SORD performance (Fig. 2d). Relative to random partitioning, the classification accuracy on MNIST, Fashion-MNIST and CIFAR-10 increases from 78%, 88%, and 67% (Fig. 2e) to 94%, 94%, and 70%, respectively (Fig. 2f). Notably, this performance advantage persists as the task complexity increases, maintaining strong accuracy even under three- and four-way (K = 3, 4) partitioning. (Supplementary Information Note 5).

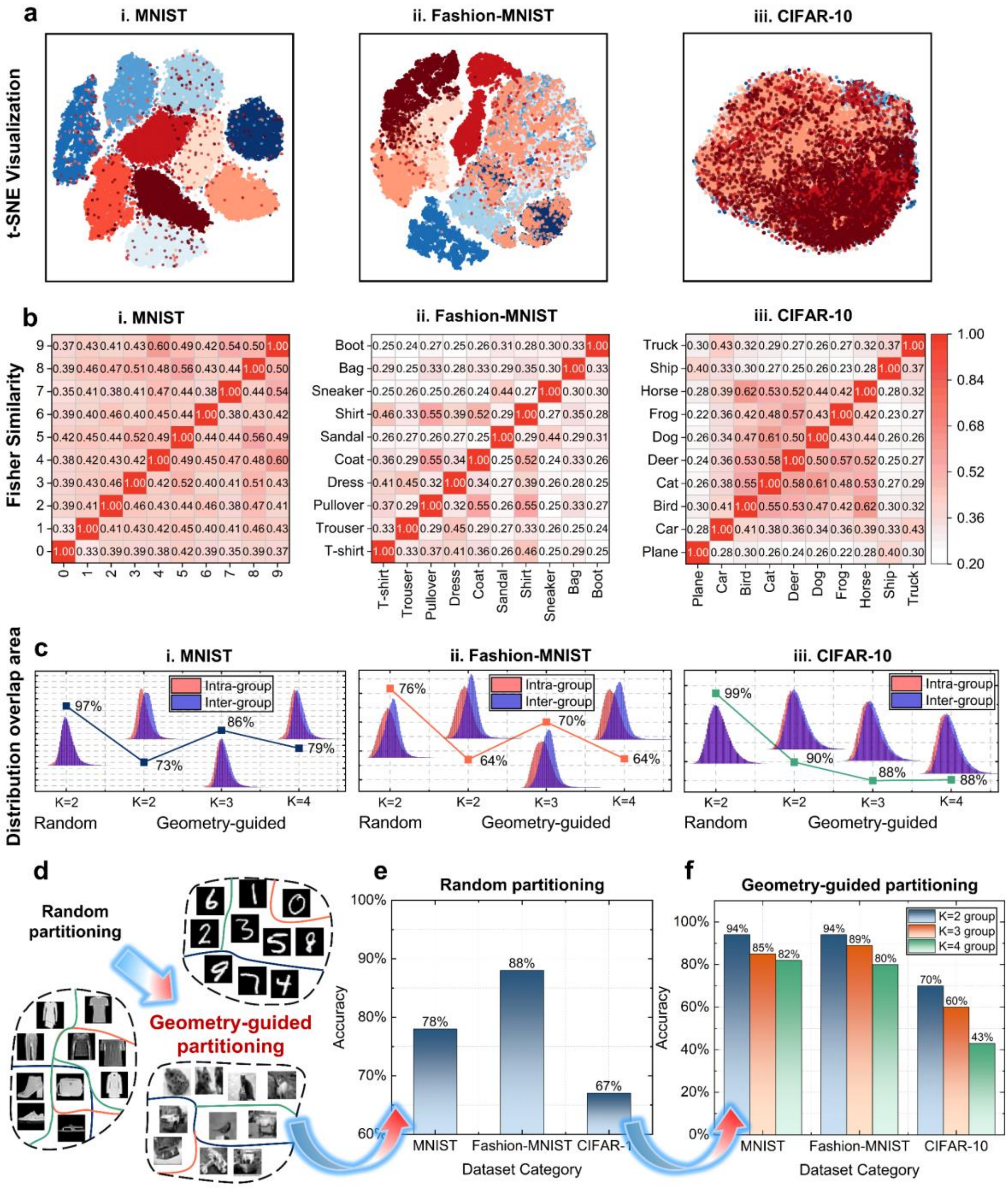

**Fig. 2 | Geometry-guided partitioning resolves entangled data manifolds. a**, t-SNE visualization of MNIST, Fashion-MNIST and CIFAR-10. **b**, Pairwise class-similarity heatmaps revealing block-diagonal structures with high-affinity clusters. **c**, Probability density estimation of intra-class (red) versus inter-class (blue) feature distances; overlap area (OA) quantifies linear inseparability. **d,** Conceptual illustration of geometry-guided grouping. **e,** Classification performance with random partitioning. **f,** Classification performance with geometry-guided partitioning, showing improved accuracy.

## SORD enhanced sensing-computing integrated system

To rigorously evaluate the performance of the proposed SORD architecture, we constructed an end-to-end optical perception system for resolving a highly entangled 100-class speckle recognition task. Speckle-based recognition represents a challenging yet fundamental problem in optical sensing, as the underlying patterns arise from high-dimensional modal interference and exhibit strong sensitivity to perturbations (Movie S1). The speckle dataset was constructed using a fiber sensor with a 40 mm twist period as the sensing front-end (Fig. 3b, Supplementary Information Note 6, 7), comprising 25 distinct spatial impact positions, each subjected to 4 discrete pressure magnitudes (0.4N, 0.8N, 1.2N and 1.6N). The pre-twisted geometry deliberately breaks the inherent rotational symmetry of the fiber, enabling a highly spatially specific mechanical response (Supplementary Information Note 8). Furthermore, to suppress potential speckle degeneracy and ensure the uniqueness of each sampling point, all pressure loading points were confined within a single twist period of the fiber. Localized pressure perturbations induce birefringence variations and strong inter-modal coupling. By directly mapping spatially varying mechanical stimuli to their corresponding speckle responses, we established a physically encoded optical dataset for evaluating hierarchical optical inference.

As shown in Fig. 3a, the optical fiber speckle field is collimated, expanded, and polarization-filtered to prepare a well-defined wavefront. A beam splitter then divides this signal into two parallel channels. The reflected path captures the ground-truth speckle patterns via a CCD camera, establishing a reliable reference. Simultaneously, the transmitted path directs the optical information straight into the physical inference module. Within this module, a SLM loaded with a pre-trained phase mask serves as the core diffractive operator (Supplementary Information Note 9). It dynamically modulates the incident complex wavefront, spatially routing the optical energy to focus onto designated detection regions on a beam profiler (Fig. 3c, Supplementary Information Note 10 and Video S2).

We adopted a physics-driven partitioning scheme governed by the natural hierarchy of the sensing process rather than the geometry-based partitioning strategy, to avoid computational overhead. As visualized by the t-SNE mapping in Fig. 3d, this physical guided partitioning yields well-clustered feature distributions, effectively enhancing local class separability. Ultimately, this approach establishes an efficient and viable pathway for practical system implementation (Supplementary Information Note 11).

The inference loop is strictly governed by a dynamic “detect-decide-refresh” protocol (Figs. 3e and 3f), where the activation of selective computational pathways and their corresponding diffractive phase masks is exclusively based on the intermediate outputs of the preceding stage.

**Stage I Macroscopic topological localization:** The SLM is initially loaded with a primary global phase mask $f_1$. This operator executes coarse-grained feature extraction on the input optical speckle to identify the generalized topological region of the mechanical stimulus, for instance the region 3. This projection effectively collapses the global search space, isolating a restricted sub-manifold $A_3$ comprising 20 candidate states.

**Stage II Precise spatial pinpointing:** Conditioned on the macroscopic verdict from layer 1 at stage I, the system retrieve a context-specific secondary mask $f_{1,3}$, from the five available alternative masks ($f_{1,j}$, $j \in [1,5]$) in the secondary layer. Upon refreshing the SLM, the system remodulates the original optical wavefront to resolve the spatial fine-structure within the isolated region. By orthogonally projecting out irrelevant broad intra-class variations, the optical field is sharply focused to pinpoint the exact millimeter-scale contact coordinate. For instance, it identifies point $P_{3,4}$ within the region $A_3$, corresponding to the 14-th position of the total 25 positions.

**Stage III Fine-grained force quantification:** Guided by the spatial coordinates confirmed in the prior stages, the system conditionally activates a corresponding tertiary mask $f_{1,3,4}$, from the third layer. This final diffractive projection resolves the residual intra-class ambiguity of the applied pressure. The definitive output $N_{3,4,3}$ represents the 3$^{rd}$ force level at the identified point $P_{3,4}$ 1.2 N, derived from the original speckle propagating through the selected 12$^{th}$ phase mask of the layer 3 library $f_{1,j,k}$, $j$, $k \in [1,5]$. Thus, the SORD effectively identifies the input stimuli from 100 entangled configurations, pinpointing a 1.2 N of pressure applied at the 14 mm axial coordinate with high fidelity. (Supplementary Information Note 12).

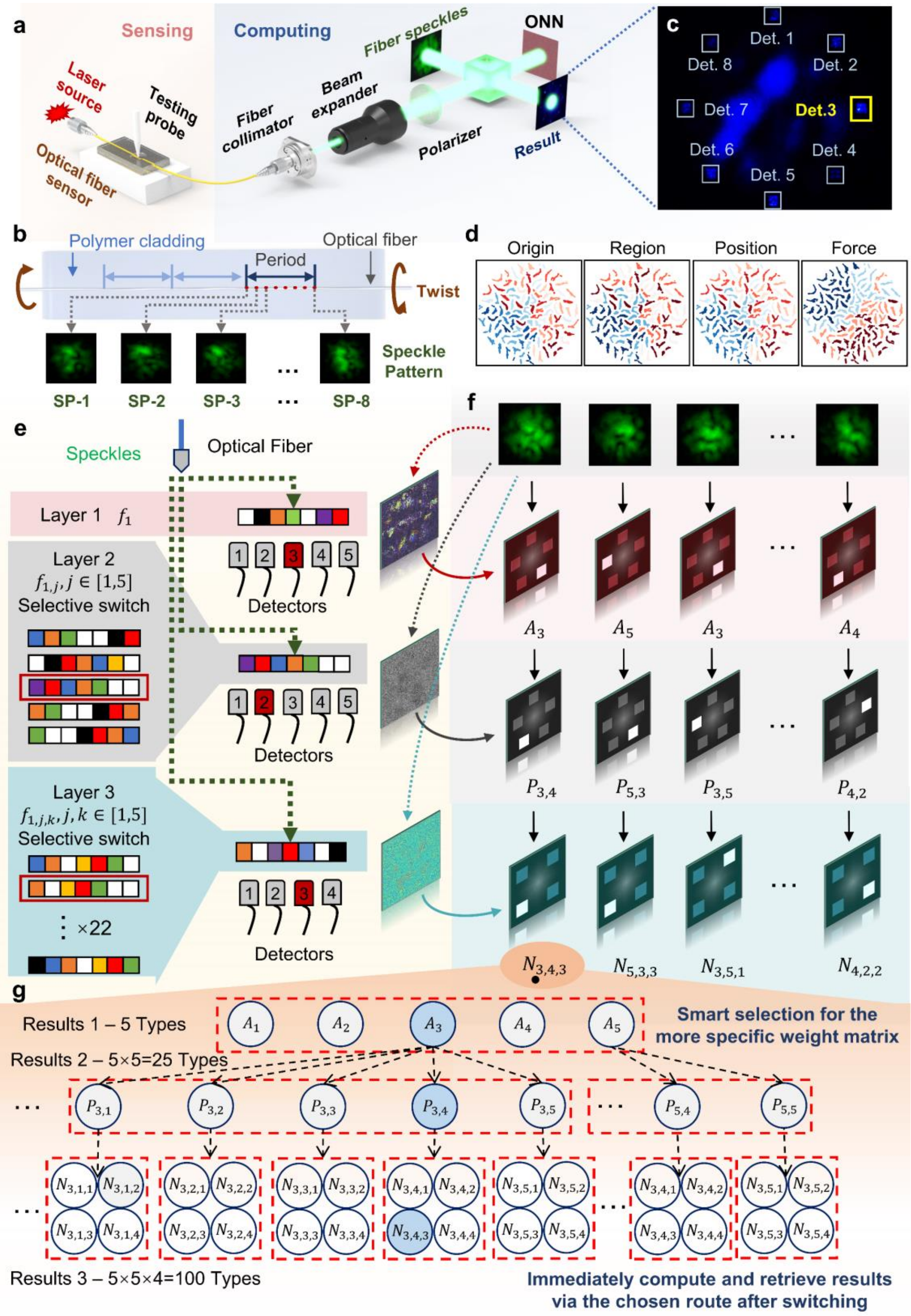

**Fig. 3 | Experimental implementation of the SORD system and hierarchical resolution of a highly entangled 100-class speckle manifold. a**, Experimental setup integrating the fiber-optic sensing

front-end and the SLM-based diffractive computing core. **b**, Schematic diagram of the polymer-encapsulated twisted fiber sensor and typical speckle patterns under external mechanical stimuli. **c**, Magnified schematic of the output plane, showing target-specific focal spots generated via ONN-mediated light-field modulation. **d**, t-SNE visualization of the 100-class speckle dataset under the original distribution and three partitioning schemes (region-, position- and force-based). **e**, Workflow of the SORD architecture for the 100-class task across three sequential stages: macroscopic topological localization, precise spatial pinpointing, and fine-grained force quantification. **f**, Dynamic phase mask switching mechanism governed by the "detect-decide-refresh" protocol, detailing the conditional activation sequence $f_1 \rightarrow f_{1,j} \rightarrow f_{1,j,k}$. **g**, Pipeline of the SORD sequential reasoning algorithm. The condition-dependent routing acts as a piecewise linear approximation, factorizing the 100-class nonlinear entanglement into a sequence of resolvable linear sub-problems.

## Empirical performance of SORD on 100-category classification and human-machine interaction

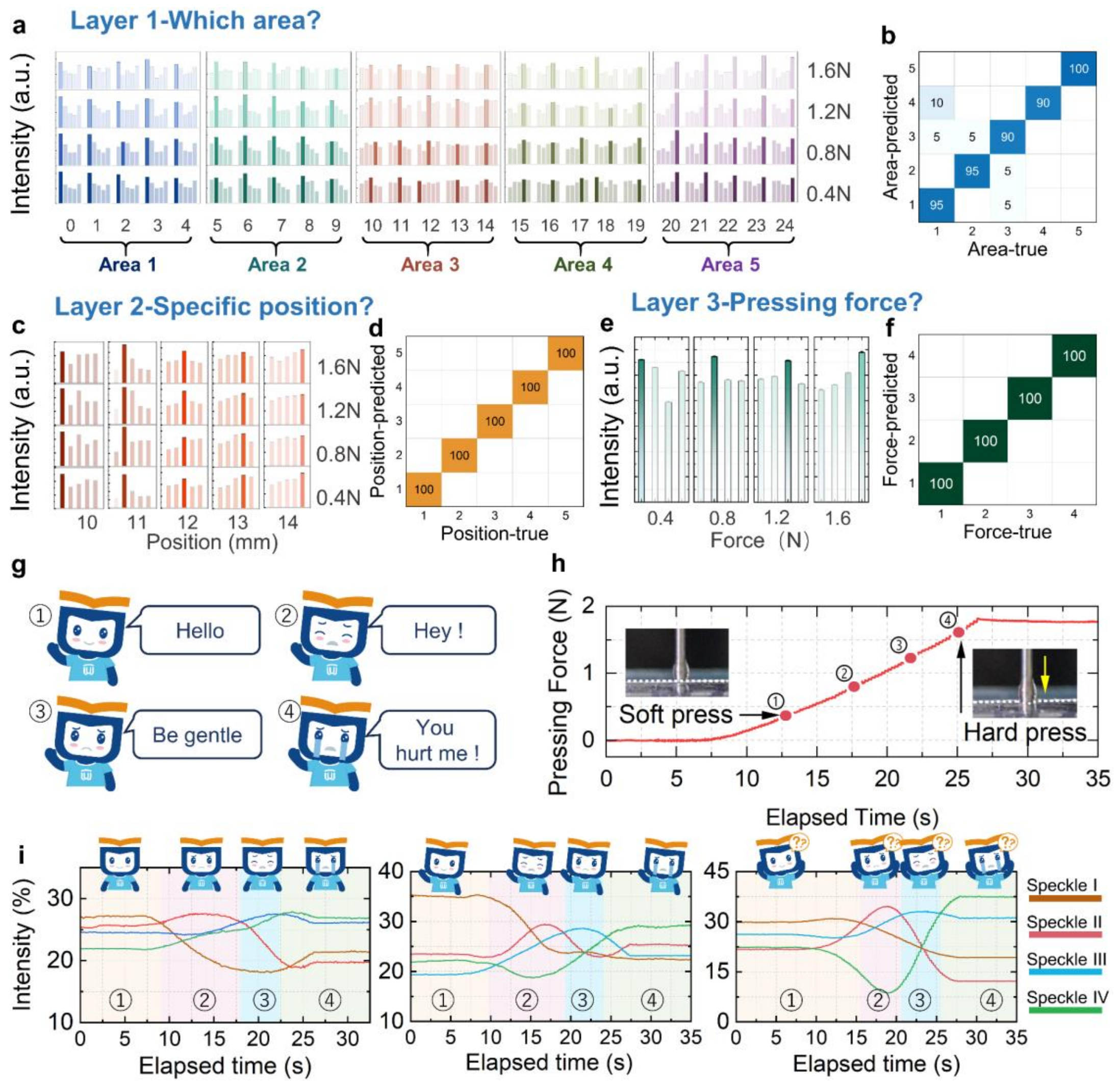

**Fig. 4 | Performance demonstration of the SORD. a**, **b**, Macroscopic region identification (Layer 1). **a**, Optical intensity distributions of 100 data points after modulation by the primary mask $f_1$ with (**b**) corresponding 5-class confusion matrix. **c**, **d**, Fine-grained spatial pinpointing within the determined region $A_3$. (Layer 2). **c**, Intensity distributions for the 20 subset data points modulated by the layer 2 mask $f_{1,3}$, with (**d**) depicting the corresponding 5-class confusion matrix. **e**, **f**, Force magnitude quantification at point $P_{3,4}$. **e**, Intensity distributions for the 4 pressure states modulated by the Layer 3 mask $f_{1,3,4}$. **f**, corresponding 4-class confusion matrix. **g**, **h**, Schematic of the human-machine interaction application. The system identifies varying applied pressures (**g**) to trigger real-time, differentiated digital facial expressions (**h**). **i**, Detailed interaction results across 12 distinct states. The rows labeled speckle 1–4 depict optical spot intensity variations corresponding to four pressure levels, evaluated under three specific interaction scenarios (different in poses corresponds in the left, middle and right panel of (**i**) respectively).

The primary diffractive operator (layer 1) partitions the 100-class speckle inputs into five regions with ~94% accuracy (Figs. 4a, b). In subsequent stages, the progressively localized sub-manifolds exhibit reduced complexity, enabling more accurate inference. For example, in the region-3 branch (20 fine-grained states), layer 2 decouples spatial coordinates from force intensities (Figs. 4c, d), identifying five impact locations with near-perfect accuracy (~100%) and strong robustness to pressure variations. Finally, operating within this collapsed spatial subspace, layer 3 resolves the residual pressure magnitude (Figs. 4e, f). As the task is progressively decomposed, the effective search space contracts and the task complexity decrease, leading to consistently improved accuracy across layers. With SORD, a single-layer diffractive optical neural network achieves 100-class classification, effectively exceeding its intrinsic capacity without additional physical resources. This highlights the power of sequential, condition-dependent inference in resolving high-complexity tasks within physically shallow optical systems.

While temporal multiplexing in SORD introduces additional latency, the resulting response time remains well within acceptable bounds for practical applications. To substantiate the practical potential of the SORD framework for edge computing, we deployed SORD in a real-time human-machine interface featuring a digital avatar "Xiao Xi" (Fig. 4g). The avatar exhibits three poses (normal, lift left hand, lift both hand) determined by the detected contacted location, and four distinct expressions (“hello”, “hey!”, “be gentle”, and “you hurt me”) governed by the magnitude of the applied force.

We use a soft force probe to apply a continuous pressure signal ramping from 0 N to 1.6 N across three distinct sites (Fig. 4h). Specifically, the applied force continuously modulates the relative optical intensities of the expression-related output spots, whose competitive redistribution determines the final facial expression of the avatar (Fig. 4i and Supplementary Information Note 14).

This implementation demonstrates that SORD enables decision-making to be directly embedded within optical signal processing. By introducing temporal sequencing, the system achieves enhanced decision capability for complex tasks while maintaining low latency operation within the bounds of real-time interaction. This establishes a practical trade-off in which temporal cost is effectively exchanged for increased computational expressivity without compromising deployment feasibility. Additionally, by unifying sensing and computation at the physical layer, this paradigm provides a scalable and energy-efficient route toward more sustainable and adaptive photonic intelligent systems.

## Discussion and Conclusion

This work introduces a sequential optical reasoning and decision (SORD) framework that reformulates inference as a state-dependent, coarse-to-fine process. By decomposing global tasks into hierarchical sub-tasks with dynamic routing, SORD effectively reduces local task complexity and enhances the representational capacity of physically shallow optical systems without increasing architectural depth.

We demonstrate that SORD enables the resolution of highly entangled real-world physical signals. In particular, the system achieves high-fidelity classification of 100 complex fiber speckle states with accuracy exceeding 90%, while supporting robust, real-time human–machine interface driven entirely by in-situ optical sensing and computing.

At the hardware level, this capability is realized through a time-multiplexed, single-layer architecture that trades temporal sequencing for functional depth. This approach significantly reduces system complexity and energy consumption, achieving a system efficiency of 23.3 TOPS/W with near real-time latency, and eliminates the need for cascaded optical layers or active nonlinear components.

Beyond the specific implementation, SORD establishes a general paradigm for integrating sensing and computation within a unified optical framework. By directly transforming complex wavefront information into structured, decision-ready representations, it provides a scalable and energy-efficient pathway toward intelligent optical systems, particularly for deployment in resource-constrained edge and wearable platforms.

**Acknowledgements**

This work was financially supported by the National Natural Science Foundation of China (Grant No. 62305309), and the Zhejiang Provincial Natural Science Foundation of China (Grant No. LMS26F050004).

# Methods

## Experimental system

The experimental setup of the proposed all-optical perception system comprises three main modules: signal generation, diffractive computing, and intensity detection. The encoded speckle field is collimated, expanded, and polarization-filtered before being divided by a beam splitter. The reflected path serves as the data acquisition channel, where a CCD camera captures fiber speckles to establish the ground truth for model training. The transmitted path constitutes the physical inference module: the optical field is projected onto a SLM loaded with a pre-trained phase mask. Functioning as the programmable diffractive operator for a specific SORD reasoning layer, the SLM modulates the incident complex speckle wavefront to focus light into designated detection regions on a beam profiler (Supplementary Information Note 10).

**a) Light source and signal coupling:** A 532 nm all-solid-state laser (Changchun New Industries, MGL-FN-532-AOM) was coupled into a single-mode fiber (Corning, SMF-28E) sensor, with the output wavefront collimated (LBTEK, FBE-3X-A) and was divided by a beam splitter (LBTEK, MBS1455-A) into two paths for ground truth acquisition and optical computing.

**b) Sensing and actuation platform:** The pressure probe was driven by a high-precision 3D motion control stage (Daheng Optics, GCD040102) to exert controlled force on the sensor, with the resulting pressure information detected by a 3D force sensor (SRI, M3815A).

**c) Ground truth acquisition:** A charge coupled device (CCD) camera (Basler, acA1920-25gc) captures the raw speckle patterns from the first path to construct the ground truth dataset for network training and validation.

**d) Optical computing Core:** A reflective SLM (UPOLabs, HDSLM80R; 1920×1200, 8 μm pitch, 8-bit depth), operated at 60 Hz, functioned as the diffractive computing unit by dynamically loading phase masks to hierarchically process the input wavefront.

**e) Result detection:** The computed optical field reflected from the SLM surface is captured by a beam quality analyzer (BeamOn U3) to record the spatial intensity distribution as the network's processing result.

## Dataset preparation

**Common parameters:** Unless otherwise specified, all datasets were collected using a twisted optical fiber sensor with a 40 mm twist period. For the optical computing experiments, the collected raw speckle images served as the amplitude input, assuming a uniform phase distribution.

**a) 2-class classification and coarse classification datasets:**

**Binary classification:** The dataset captured speckle patterns at a fixed position. Valid samples were filtered based on a signal-to-noise criterion, specifically selecting instances where the response intensity exceeded the system's 3σ mechanical noise floor.

**System performance limits benchmark:** As classification cardinality increases, the DONN exhibits clear signs of saturation, characterized by compromised robustness, decelerated convergence, and deteriorated stability. To probe the topological performance limit of single-layer diffractive structures for better delineation of application boundaries, a low-sensitivity fiber sensor (40 mm twist period) is employed to construct a rigorous and challenging worst-case scenario benchmark. As shown in Supplementary Fig. S4, recognition accuracy suffers a precipitous decline as speckle similarity (SSIM) increases. The simulated classification accuracy falls below 80% when the SSIM exceeds 0.97, and the experimentally intensity ratio between the target and the competing spot approaches parity (1:1). This failure physically corroborates that a global linear diffractive system cannot effectively segregate samples that are highly overlapped in the high-dimensional input domain, confirming a definitive linear separability ceiling for the single-layer diffractive networks.

**4-class and 8-class tasks:** For coarse classification benchmarks, data were collected by applying pressing force across discrete positions. These datasets comprise 500 samples per class, yielding total sizes of 2,000 and 4,000 samples, respectively. Supplementary Figs. S16 and S17 illustrate the convergence profiles of the model under varying hyperparameters in theory; the model achieves validation accuracies of 100% and >95% for 4-class and 8-class classification tasks, respectively (Supplementary Information Note 13). Remarkably, in the physical optical implementation, leveraging the enhanced generalization capability, the system demonstrated superior performance, attaining near-perfect (~100%) classification accuracy in both 4-class and 8-class tasks. This experimental performance indicates that the trained model has successfully learned invariant physical features robust to hardware perturbations, effectively bridging the gap between theoretical simulation and optical reality.

**b) 100-class fine-grained dataset:** To validate the practical efficiency of the SORD architecture, we constructed a comprehensive dataset encompassing 100 distinct mechanical stimulation states, using a fiber sensor with a 40 mm twist period as the sensing front-end. This specific configuration was chosen not only to accommodate the extensive 25-point spatial range without periodic ambiguity, but also to maximize signal correlation between adjacent classes. The resulting high-similarity feature manifolds present a stringent challenge that exceeds the discriminative capacity of conventional ONN. The data acquisition involved a pressure probe translating along the fiber axis at 1 mm intervals across 25 spatial points, with four discrete pressure magnitudes 0.4N, 0.8N, 1.2N and 1.6N applied at each location.

This dataset covered a 25 mm sensing region sampled at 1 mm spatial intervals. At each of the 25 positions, 4 distinct levels (0.4 N, 0.8 N, 1.2 N, 1.6 N were applied, resulting in 100 distinct classes (25 positions × 4 forces). To support the multi-stage ONN architecture, the data is structured as follows:

**Layer 1** utilizes 50,000 samples (5 super-classes×10,000 samples per class);

**Layer 2** contains 10,000 samples (5 positions×2,000 samples per class);

**Layer 3** contains 2,000 samples (4 forces×500 samples per class).

**c) HMI dataset:** For the HMI demonstration, a specialized twisted optical fiber sensor with a 20 mm periodic was employed. Speckle data were acquired at three discrete spatial coordinates (x=5 mm, x=10 mm, and x=15mm) along the fiber. At each location, mechanical stimuli were applied at four distinct force levels (0.4 N, 0.8 N, 1.2 N, 1.6 N).

**d) Standard computer vision dataset:**

·MNIST dataset contains 10-category handwritten digits, comprising 60,000 training samples and 10,000 testing samples[53].

·Fashion-MNIST dataset is a collection of fashion products of 10 different categories, also containing 60,000 samples and 10,000 testing samples[54].

·CIFAR-10 dataset is a subset of the 80 million tiny images dataset, consisting 50,000 training images and 10,000 testing images[55].

## The SORD learning procedure

To address large-scale and complex classification challenges, the SORD framework operates through a cyclical four-stage process: (i) Task decomposition, (ii) Optical propagation and computing, (iii) Detection and selective switching, and (iv) Recursive circulation until the final output is generated.

The SORD architecture effectively transcends the capacity limitations of conventional single-layer ONNs. Rather than forcing a single mathematically impossible linear separation on a highly entangled input distribution, SORD strategically decomposes the global complex problem into a sequence of low-entropy, condition-dependent sub-tasks. Fundamentally, by relying on progressive manifold partitioning through iterative logical decisions, the architecture isolates and linearizes local feature spaces. Consequently, the ultimate solving capability of the entire system for hyper-complex tasks is effectively lower-bounded by its baseline physical proficiency in resolving a fundamental 2-class binary classification within a highly restricted subspace. To validate this universal computing framework and the versatility of the manifold partition strategy, the standard computer vision datasets MNIST/Fashion-MNIST/CIFAR-10 were partitioned into 2/3/4-class group based on Fisher score. All combinations were classified only using the first layer to simulate the classification performance at the first step, which undertakes the most arduous task for the most data and global manifold and whose accuracy can be a key representative of performance improvement. The high classification accuracy verifies the universality of the proposed computing framework. This implies that, theoretically, as long as the binary classification accuracy is sufficient, the system can scale to handle tasks of increasing complexity.

## Neural network configuration

**a) Physical-digital mapping:** The diffractive neural network is constructed based on a forward propagation model. The SLM utilizes an 8-bit depth, where the 0-255 grayscale levels are linearly mapped to a phase modulation range of 0-2π. To fully exploit the modulation bandwidth of the SLM, the incident optical speckle was expanded to a diameter of approximately 9 mm via the beam expansion optics. Accordingly, the physical dimension of the diffractive neural network layer was defined as 9.6×9.6 mm to encompass the optical field. The input images were zero-padded to 1000×1000 pixels to ensure precise alignment and prevent boundary artifacts during the free-space diffraction simulation.

**b) Diffractive geometry and model optimization:** The effective optical path lengths were calibrated to 17.5 cm (fiber output to SLM) and 22.5 cm (SLM to detector). These distances were optimized to satisfy the sampling criteria for free-space diffraction simulation, while ensuring the generation of robust, high-contrast interference patterns at the detection plane. Utilizing this physical propagation theory, the network parameters were subsequently determined and optimized via simulation training, configured to map the optical intensity concentration at predefined detection regions corresponding to the ground-truth class labels.

## Optical Neural network training

**a) Computational environment and optimization:** All training procedure were implemented using the PyTorch framework (version 2.2.0) with Python (version 3.9.20). Computations were accelerated by an NVIDIA A100 GPU (80 GB), a CPU with 16 cores and unlimited RAM size in the Linux system hosted at the High-Performance Computing Center of Westlake University. The network parameters—specifically the phase mask weight matrices—were optimized using the adaptive moment estimation (Adam) algorithm[56]. The objective was to minimize the mean squared error (MSE) loss between the detected optical intensity distribution and the ground-truth labels[57, 58].

**b) Hierarchical training protocol:** While the standard classification tasks (2-, 4-, and 8-class) were achieved using a single ONN layer which converged within 50 epochs, the complex tasks utilized the proposed SORD architecture to decompose the problem space:

**1) 100-Class task:** This task was trained across three sequential layers. Layer-1 classified inputs into 5 coarse regions (each encompassing 20 fine-grained classes); Layer-2 identifies the specific spatial position within the determined region (5 sub-positions); and Layer-3 resolves the applied force magnitude (4 levels).

**2) HMI task:** The 12-state HMI task was decomposed into two layers. Layer-1 performed spatial localization to identify the active touching region (3 areas), while Layer-2 performed force quantification (4 pressure levels) for the identified area.

**c) Robustness strategies and training settings:** To counteract systemic errors inherent in the experimental setup such as wavefront distortions, beam size mismatches, and SLM misalignments, physics-aware data augmentation techniques were applied, including random rotation, affine transformation, elastic transformation, perspective distortion, and intensity jitter. The network was trained using a custom loss function with a learning rate of 0.001. The batch size was set to 64 for standard tasks. For complex hierarchical tasks, dynamic batch sizes were employed: 256, 128, and 64 for layers 1, 2, and 3 of the 100-class tasks, respectively; and 128 and 64 for layers 1 and 2 of the HMI task. Both hierarchical models also trained over 50 epochs per layer for convergence. All networks were trained using the full dataset, partitioned into an 80:20 split for training and validation, respectively, before final deployment and testing on the experimental optical system.

## System computing performance and benchmarking

The performance benchmarks of the proposed perception system, compared against conventional perception strategies and existing ONNs, are detailed in Supplementary Information Tables S1, respectively. The calculation methodologies are provided in the Supplementary Information Note 15.

**a) Computing speed:** The computational throughput is quantified by the total number of operations performed per second. For the single-layer DONN utilized in this work, a single inference pass performs 42.32 tera operations (OPs). Leveraging the dynamic modulation capability of the SLM operating at 60 Hz, the system achieves a peak computing speed of 2,539.2 tera operations per second (TOPS).

**b) System energy efficiency:** We adopted a stringent system-level metric to evaluate energy efficiency, accounting for the wall-plug power consumption of all active components, including the desktop computer interface, the SLM, and the solid-state laser (Supplementary Information Table S2). With a total measured power consumption of approximately 43 W, the prototype achieves a system energy efficiency of 23.3 TOPS/W, demonstrating that the proposed SORD architecture allows for scalable computing depth to handle complex perception tasks without increasing hardware complexity.